\documentclass[aps,prl,preprint,tightenlines,superscriptaddress,showpacs,byrevtex]{revtex4}
\usepackage{graphicx}
\usepackage{epsfig}
\usepackage{dcolumn}
\usepackage{bm}
\usepackage{color}

\newcommand{\MMS}{M_{\rm rec}^2}
\newcommand{\y}{Y(4260)}
\newcommand{\lum}{{\cal L}}
\newcommand{\eff}{\varepsilon}
\newcommand{\BR}{{\cal B}}
\newcommand{\jpc}{J^{PC}}
\newcommand{\pip}{\pi^+}
\newcommand{\pim}{\pi^-}
\newcommand{\piz}{\pi^0}

\newcommand{\psp}{\psi(2S)}
\newcommand{\pspp}{\psi(3770)}
\newcommand{\jpsi}{J/\psi}
\newcommand{\psift}{\psi(4040)}
\newcommand{\psifto}{\psi(4160)}
\newcommand{\psiftf}{\psi(4415)}
\newcommand{\EE}{e^+e^-}
\newcommand{\MM}{\mu^+\mu^-}
\newcommand{\LL}{\ell^+\ell^-}
\newcommand{\pp}{\pi^+\pi^-}
\newcommand{\kk}{K^+K^-}
\newcommand{\ddb}{D\overline{D}}

\newcommand{\ppjpsi}{\pi^+\pi^- J/\psi}
\newcommand{\jpsipp}{\pi^+\pi^- J/\psi}
\newcommand{\beq}{\begin{equation}}
\newcommand{\eeq}{\end{equation}}
\newcommand{\bitm}{\begin{itemize}}
\newcommand{\eitm}{\end{itemize}}

\def\Journal#1#2#3#4{{#1} {\bf #2}, #3 (#4)}

\def\PLB{Phys. Lett. B}
\def\PRL{Phys. Rev. Lett.}
\def\PRD{Phys. Rev. D}

\begin{document}

\preprint{} \preprint{\vbox{ \hbox{   }
                        \hbox{Belle Preprint 2007-31}
                        \hbox{KEK   Preprint 2007-23}
                        }}
\title{\quad\\[1.0cm]
Measurement of $\EE \to \ppjpsi$ Cross Section via Initial State
Radiation at Belle}

\affiliation{Budker Institute of Nuclear Physics, Novosibirsk}
\affiliation{Chiba University, Chiba} \affiliation{University of
Cincinnati, Cincinnati, Ohio 45221} \affiliation{Department of
Physics, Fu Jen Catholic University, Taipei}
\affiliation{The Graduate University for Advanced Studies, Hayama}
\affiliation{Hanyang University, Seoul} \affiliation{University of
Hawaii, Honolulu, Hawaii 96822} \affiliation{High Energy
Accelerator Research Organization (KEK), Tsukuba}
\affiliation{Hiroshima Institute of Technology, Hiroshima}
\affiliation{University of Illinois at Urbana-Champaign, Urbana,
Illinois 61801} \affiliation{Institute of High Energy Physics,
Chinese Academy of Sciences, Beijing} \affiliation{Institute of
High Energy Physics, Vienna} \affiliation{Institute of High Energy
Physics, Protvino} \affiliation{Institute for Theoretical and
Experimental Physics, Moscow} \affiliation{J. Stefan Institute,
Ljubljana} \affiliation{Kanagawa University, Yokohama}
\affiliation{Korea University, Seoul}
\affiliation{Kyungpook National University, Taegu}
\affiliation{Swiss Federal Institute of Technology of Lausanne,
EPFL, Lausanne} \affiliation{University of Ljubljana, Ljubljana}
\affiliation{University of Maribor, Maribor}
\affiliation{University of Melbourne, School of Physics, Victoria
3010} \affiliation{Nagoya University, Nagoya} \affiliation{Nara
Women's University, Nara} \affiliation{National Central
University, Chung-li} \affiliation{National United University,
Miao Li} \affiliation{Department of Physics, National Taiwan
University, Taipei} \affiliation{H. Niewodniczanski Institute of
Nuclear Physics, Krakow} \affiliation{Nippon Dental University,
Niigata} \affiliation{Niigata University, Niigata}
\affiliation{Osaka City University, Osaka} \affiliation{Osaka
University, Osaka} \affiliation{Panjab University, Chandigarh}
\affiliation{Princeton University, Princeton, New Jersey 08544}
\affiliation{RIKEN BNL Research Center, Upton, New York 11973}
\affiliation{University of Science and Technology of China, Hefei}
\affiliation{Seoul National University, Seoul}
\affiliation{Sungkyunkwan University, Suwon}
\affiliation{University of Sydney, Sydney, New South Wales}
\affiliation{Toho University, Funabashi} \affiliation{Tohoku
Gakuin University, Tagajo} \affiliation{Tohoku University, Sendai}
\affiliation{Department of Physics, University of Tokyo, Tokyo}
\affiliation{Tokyo Metropolitan University, Tokyo}
\affiliation{Tokyo University of Agriculture and Technology,
Tokyo}
\affiliation{Virginia Polytechnic Institute and State University,
Blacksburg, Virginia 24061} \affiliation{Yonsei University, Seoul}
  \author{C.~Z.~Yuan}\affiliation{Institute of High Energy Physics, Chinese Academy of Sciences, Beijing} 
  \author{C.~P.~Shen}\affiliation{Institute of High Energy Physics, Chinese Academy of Sciences, Beijing} 
  \author{P.~Wang}\affiliation{Institute of High Energy Physics, Chinese Academy of Sciences, Beijing} 
  \author{S.~McOnie}\affiliation{University of Sydney, Sydney, New South Wales} 
  \author{I.~Adachi}\affiliation{High Energy Accelerator Research Organization (KEK), Tsukuba} 
  \author{H.~Aihara}\affiliation{Department of Physics, University of Tokyo, Tokyo} 
  \author{V.~Aulchenko}\affiliation{Budker Institute of Nuclear Physics, Novosibirsk} 
  \author{T.~Aushev}\affiliation{Swiss Federal Institute of Technology of Lausanne, EPFL, Lausanne}\affiliation{Institute for Theoretical and Experimental Physics, Moscow} 
  \author{S.~Bahinipati}\affiliation{University of Cincinnati, Cincinnati, Ohio 45221} 
  \author{V.~Balagura}\affiliation{Institute for Theoretical and Experimental Physics, Moscow} 
% ...
\collaboration{The Belle Collaboration}

\date{\today}

\begin{abstract}

The cross section for $\EE\to \ppjpsi$ between 3.8 and
5.5~GeV/$c^2$ is measured using a 548~fb$^{-1}$ data sample
collected on or near the $\Upsilon(4S)$ resonance with the Belle
detector at KEKB. A peak near 4.25~GeV/$c^2$, corresponding to the
so called $\y$, is observed. In addition, there is another cluster
of events at around 4.05~GeV/$c^2$. A fit using two interfering
Breit-Wigner shapes describes the data better than one that uses
only the $\y$, especially for the lower mass side of the 4.25~GeV
enhancement.

\end{abstract}


\maketitle

In a recent study of initial state radiation ($ISR$) events of the
type, $\EE \to \gamma_{ISR} \ppjpsi$, the BaBar Collaboration
observed an accumulation of events near 4.26~GeV/$c^2$ in the
$\ppjpsi$ invariant mass distribution and attributed it to a
possible new resonance that they dubbed the $\y$~\cite{babay4260}.
This observation was confirmed by the CLEO experiment using a
similar technique with a data sample collected at the
$\Upsilon(4S)$ peak~\cite{cleo_y}. The CLEO Collaboration also
collected a 13.2~pb$^{-1}$ data sample at $\sqrt{s}=4.26$~GeV, and
reported signals for $\ppjpsi$, $\piz \piz \jpsi$, and $\kk \jpsi$
with cross sections that are significantly higher than those
measured at other nearby energies~\cite{cleoy4260}.

Since the $\y$ resonance is produced via $\EE$ annihilation
accompanied with initial state radiation, its $\jpc=1^{--}$.
However, the properties of the observed peak are rather different
from those of other known $\jpc=1^{--}$ charmonium states in the
same mass range, such as $\psift$, $\psifto$, and $\psiftf$. Since
it is well above the $\ddb$ threshold, it is expected to decay
predominantly into $D^{(*)} \bar{D}^{(*)}$ final states. The
partial width for the $\pi\pi\jpsi$ final state is expected to be
a small fraction of the total.  In fact, the $\y$ shows an
unusually strong coupling to the $\pi\pi\jpsi$ final state while
no significant enhancement is observed in $D^{(*)} \bar{D}^{(*)}$
final states~\cite{pakhlova}. In a fit to the total hadronic cross
sections measured by the BES experiment~\cite{besr1,besr2} for
$\sqrt{s}$ between 3.7 and 5.0~GeV, Mo {\em et al.} set an upper
limit on $\Gamma_{\EE}$ for the $\y$ to be less than 580~eV at
90\% confidence level (C.L.)~\cite{moxh_y}. This implies that its
branching fraction to $\pi\pi\jpsi$ is greater than 1.3\% at 90\%
C.L. These properties have triggered many models to explain the
$\y$ as an exotic state, such as a four-quark state, a molecular
state, or a quark-gluon hybrid~\cite{swanson}.

In the analysis reported here, we use a 548~fb$^{-1}$ data sample
collected with the Belle detector~\cite{Belle} operating at the
KEKB asymmetric-energy $e^+e^-$ (3.5 on 8~GeV)
collider~\cite{KEKB} to investigate the $\ppjpsi$ final state
produced via $ISR$. About 90\% of the data were collected at the
$\Upsilon(4S)$ resonance ($\sqrt{s}=10.58$~GeV), and about 10\%
were taken at a center-of-mass (CM) energy that is 60~MeV below
the $\Upsilon(4S)$ peak. The measurement in this Letter uses an
improved efficiency for detecting $ISR$ events, and supersedes the
preliminary results in Ref.~\cite{Belle_y}, which confirmed the
structure near 4.26~GeV/$c^2$.

For Monte Carlo (MC) simulations of the $ISR$ process, we generate
signal events with the PHOKHARA program~\cite{phokhara}. In this
program, after one or two photons are emitted, the lower energy
$\EE$ pair forms a resonance $X$ that  subsequently decays to
$\pip\pim\jpsi$ with the $\jpsi$ decaying either to $\EE$ or
$\MM$. In the $X\to \pip\pim \jpsi$ generation, we use pure
$S$-waves between the $\pi\pi$ system and the $\jpsi$, as well as
between the $\pip$ and $\pim$; this is in agreement with the
experimental results~\cite{besdist,babay4260}. The $\pp$ invariant
mass distributions are generated according to phase space. For
$\psp\to \ppjpsi$, which we use as a calibration process, we use
the decay properties that have been measured  with high
precision~\cite{besdist}.

For candidate events, we require the number of charged  tracks to
be four and net charge to be zero. For these tracks, the impact
parameters perpendicular to and along the beam direction with
respect to the interaction point are required to be less than
$0.5$ and $4$~cm, respectively, and transverse momentum is
restricted to be higher than 0.1~GeV/$c$.  For each charged track,
information from different detector subsystems is combined to form
a likelihood for each particle species ($i$),
$\mathcal{L}_i$~\cite{pid}. Tracks with
$\mathcal{R}_K=\frac{\mathcal{L}_K}{\mathcal{L}_K+\mathcal{L}_\pi}<0.4$
are identified as pions with an efficiency of about 95\% for the
tracks of interest. Similar likelihood ratios are formed for
electron and muon identification. For electrons from $\jpsi\to
\EE$, one track should have $\mathcal{R}_e>0.95$ and the other
$\mathcal{R}_e>0.05$; for muons from $\jpsi\to \MM$, at least one
track is required to have $\mathcal{R}_\mu>0.95$; in cases where
one of the tracks has no muon identification (ID) information, the
polar angles of the two muon tracks in the $\pip\pim \MM$ CM
system are required to satisfy $|\cos\theta_\mu|<0.7$ based on a
comparison between data and MC simulation. Lepton ID efficiency is
about 90\% for $\jpsi\to \EE$ and 87\% for $\jpsi\to \MM$. Events
with $\gamma$-conversions are removed by requiring $\mathcal{R}_e
< 0.75$ for the $\pip\pim$ tracks. For the $\jpsi\to \EE$ mode,
$\gamma$-conversion events are further removed by requiring the
$\pip\pim$ invariant mass to be greater than 0.35~GeV/$c^2$.

The detection of the $ISR$ photon is not required, instead, we
identify $ISR$ events by the requirement
$|\MMS|<2.0~(\hbox{GeV}/c^2)^2$, where $\MMS$ is the square of the
mass that is recoiling against the four charged tracks.

Clear $\jpsi$ signals are observed in both decay modes. We define
a $\jpsi$ signal region  as $3.06~{\rm GeV}/c^2 < m_{\ell^+\ell^-}
< 3.14~{\rm GeV}/c^2$ (the mass resolution is about 17~MeV/$c^2$),
and $\jpsi$ mass sidebands as $m_{\ell^+\ell^-}\in [2.91,
3.03]$~GeV/$c^2$ or $m_{\ell^+\ell^-}\in [3.17, 3.29]$~GeV/$c^2$;
the latter are three times as wide as the signal region.

\begin{figure}[htbp]
\psfig{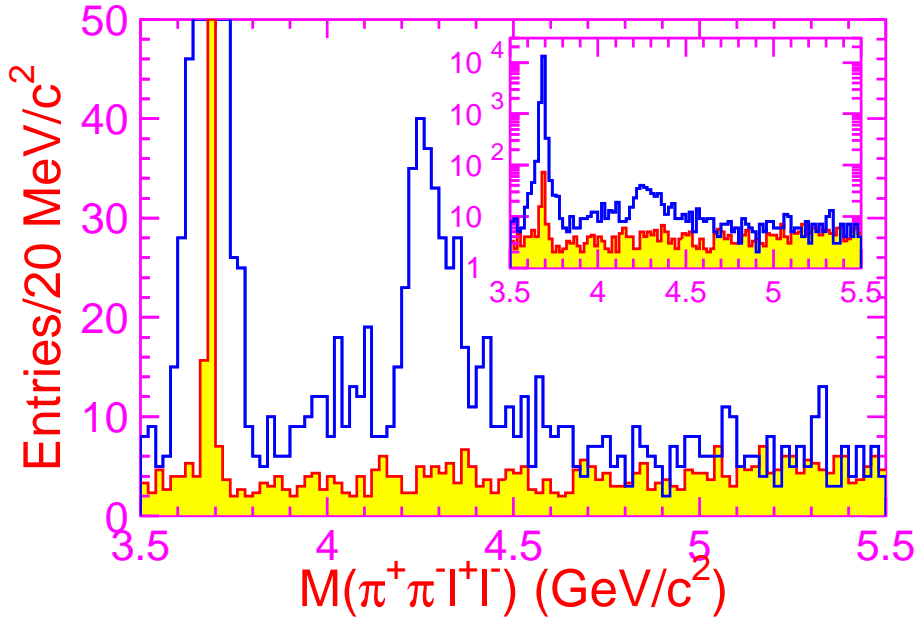}
\caption{Invariant mass distribution of $\pp\LL$. The blank
histograms represent the selected data and the shaded histograms
are the normalized sidebands. The inset shows the distribution
with a logarithmic vertical scale.} \label{mppll_full}
\end{figure}

Figure~\ref{mppll_full} shows the $\pp\LL$ invariant
mass~\cite{footnote} distribution after the above selection,
together with the background estimated from the $\jpsi$ mass
sidebands. In addition to a huge $\psp$ signal, there is a clear
enhancement at 4.25~GeV/$c^2$ similar to that observed by the
BaBar Collaboration~\cite{babay4260}. In addition, there is a
clustering of events around 4.05~GeV/$c^2$ that is significantly
above the background level. It is evident in the figure that the
background estimated from the $\jpsi$ sidebands agrees well with
the level of the selected events in the high $\pp\LL$ invariant
mass region. A study of events in the
$|\MMS|>1~(\hbox{GeV}/c^2)^2$ region, which is depleted in signal
events, supports this conclusion. The backgrounds not in the
sidebands, including: (1) $\pip\pim\jpsi$, with $\jpsi$ decays
into final states other than lepton pairs; (2) $X\jpsi$, with $X$
not being $\pip\pim$, such as $K^+K^-$ and $\pip\pim\piz$, are
found from MC simulation to be less than one event per
20~MeV/$c^2$ bin at 90\% C.L. according to the CLEO
measurements~\cite{cleoy4260} and are neglected. The production of
$\ppjpsi$ from non-$ISR$ processes, such as $\EE\to \gamma
\gamma^* \gamma^*\to \gamma \rho^0 \jpsi$, is computed to be
small~\cite{davier} and is neglected.

\begin{figure}[htbp]
\psfig{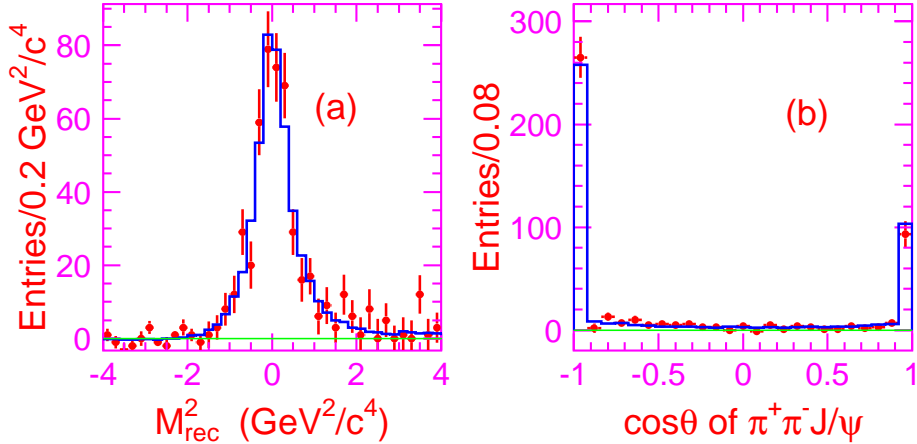}
\caption{$\MMS$ distribution (a) and the polar angle distribution
of the $\ppjpsi$ system in the $\EE$ CM frame (b) for the selected
$\ppjpsi$ events with invariant masses between 3.8 and
4.6~GeV/$c^2$. The background from $\jpsi$ mass sidebands has been
subtracted, and the selection criterion applied to the $\MMS$ has
been relaxed in (a). The points with error bars are data, compared
with MC simulation (solid histograms).} \label{recmx_xcosthe_y}
\end{figure}

The data points in Figs.~\ref{recmx_xcosthe_y}(a)~and~(b) show the
background-subtracted $\MMS$ distribution and the polar angle
distribution of the $\ppjpsi$ system in the $\EE$ CM system for
the selected $\ppjpsi$ events with invariant mass between 3.8 and
4.6~GeV/$c^2$. The data agree well with the MC simulation,
indicating that the signal events are produced via $ISR$.

We estimate the signal significance of the clusters at
4.05~GeV/$c^2$ and 4.25~GeV/$c^2$ by comparing the numbers of
signal events (number of observed events in the $\jpsi$ signal
window minus the number of $\jpsi$-sideband-estimated background
events) with their statistical uncertainties. For events with
$m_{\pp\LL}\in [3.80,4.15]$~GeV/$c^2$, we have \( n^{\rm
sig}(4.05) = 120\pm 14\), which is more than 8$\sigma$ from zero
assuming a Gaussian error; while for events with $m_{\pp\LL}\in
[4.15,4.60]$~GeV/$c^2$, we have \( n^{\rm sig}(4.25) = 324\pm 21
\), which is more than 15$\sigma$ from zero.

\begin{figure}[htbp]
\psfig{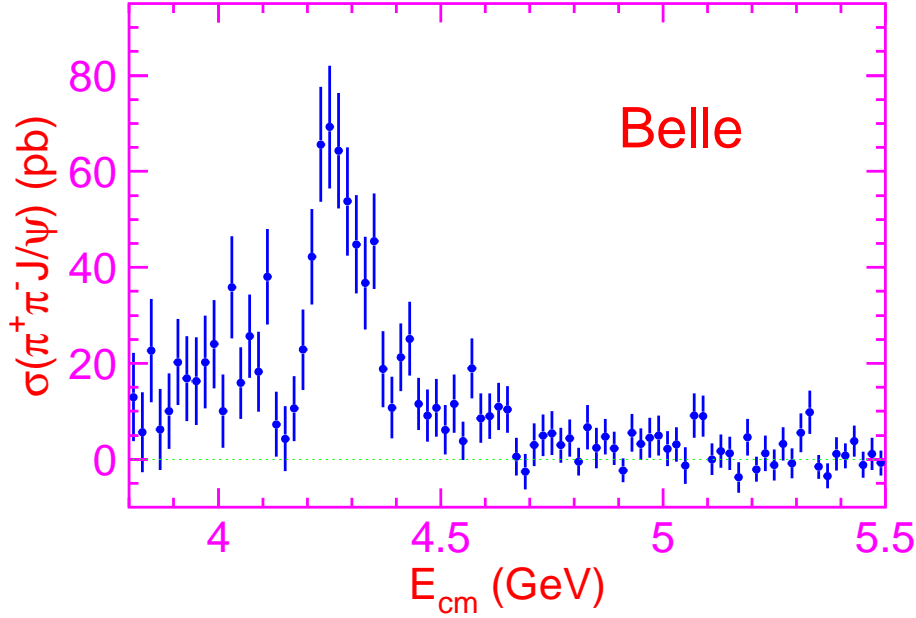} \caption{The
measured $\EE\to \ppjpsi$ cross section for CM energies between
3.8 and 5.5~GeV. The errors are statistical only.} \label{xs_full}
\end{figure}

The $\EE\to \jpsipp$ cross section for each $\ppjpsi$ mass bin
is computed with
 \(
 \sigma_i = \frac{n^{\rm obs}_i - n^{\rm bkg}_i}
                 {\eff_i \lum_i \BR(\jpsi\to \LL)},
 \)
where $n^{\rm obs}_i$, $n^{\rm bkg}_i$, $\eff_i$, and $\lum_i$ are
the number of events observed in data, the number of background
events determined from the $\jpsi$ sidebands, the efficiency, and
the effective luminosity~\cite{kuraev} in the $i$-th $\ppjpsi$
mass bin, respectively; $\BR(\jpsi\to \LL)=11.87\%$ is taken from
Ref.~\cite{PDG}. The resulting cross sections are shown in
Fig.~\ref{xs_full}, where the error bars indicate the combined
statistical errors of the signal plus background events. Our
measurement at 4.26~GeV/$c^2$ agrees well with BaBar's and CLEO's
results~\cite{babay4260,cleoy4260}.

\begin{figure}[htbp]
\psfig{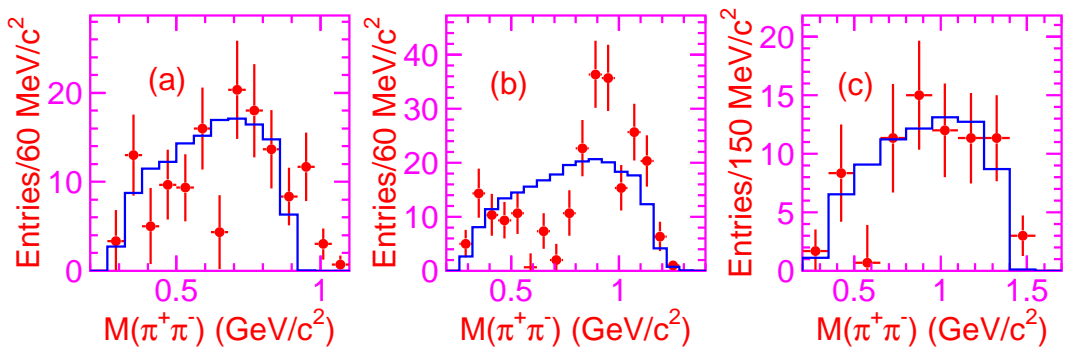}
\caption{The $\pp$ invariant mass distribution of events for
different $\ppjpsi$ mass regions. (a): $m_{\ppjpsi}\in
[3.8,4.2]$~GeV/$c^2$, (b): $m_{\ppjpsi}\in [4.2,4.4]$~GeV/$c^2$,
and (c): $m_{\ppjpsi}\in [4.4,4.6]$~GeV/$c^2$. The points with
errors bars are pure signal events, the histograms are MC
simulations made using phase space distributions.}
\label{mpipi_signal}
\end{figure}

The sources of the systematic errors for the cross section
measurement are listed in Table~\ref{err_full}. The particle ID
uncertainty, measured using the $\psp$ events in the same data
sample, is 3.0\%; the uncertainty in the tracking efficiency for
tracks with angles and momenta characteristic of signal events is
about 1\%/track, and is additive;  efficiency uncertainties
associated with the $\jpsi$ mass and $\MMS$ requirements are also
determined from a study of the very pure $\psp\to \ppjpsi$ event
sample.  In this study we find that the detection efficiency is
lower than that inferred from the MC simulation by $(2.5\pm
0.4)\%$. A correction factor is applied to the final results and
0.4\% is included in the systematic error. Belle measures the
luminosity with a precision of 1.4\% using wide angle Bhabha
events, and the uncertainty of the $ISR$ photon radiator is
0.1\%~\cite{kuraev}. The main uncertainty of the
PHOKHARA~\cite{phokhara} generator is due to the modelling of the
$\pp$ mass spectrum. Figure~\ref{mpipi_signal} shows the $\pp$
invariant mass distributions of events for three $m_{\ppjpsi}$
regions, $[3.8,4.2]$, $[4.2,4.4]$, and $[4.4,4.6]$ (unit in
GeV/$c^2$). The $\pp$ invariant mass distribution for events
around 4.25~GeV/$c^2$ differs significantly from phase space; for
other energy ranges the agreement with phase space is better.
Simulations with modified $\pp$ invariant mass distributions yield
efficiencies that are higher by 2-5\% for $m_{\ppjpsi}$ below
4.4~GeV/$c^2$. This is not corrected for in the analysis, but is
taken as the systematic error (conservatively assigned as 5\%) for
all $\ppjpsi$ mass values.  The selected events have four charged
tracks and 16-25\% of them have a detected high energy $ISR$
photon. According to the MC simulation, the trigger efficiency for
these events is around 98\%, with an uncertainty that is smaller
than 1\%. The uncertainty of $\BR(\jpsi\to \LL)=\BR(\jpsi\to
\EE)+\BR(\jpsi\to \MM)$ is taken as 1\% by linearly adding the
errors of the world averages for the $\EE$ and $\MM$
modes~\cite{PDG}. Finally the MC statistical error on the
efficiency is 1\%. We assume all the sources are independent and
add them in quadrature, resulting in a total systematic error on
the cross section of 7.5\%.

\begin{table}[htbp]
\caption{Systematic errors in the cross section measurement. They
are common for all data points.} \label{err_full}
\begin{center}
\begin{tabular}{c | c}
\hline
  Source & Relative error (\%) \\\hline
 Particle ID &  3.0 \\
 Tracking & 4 \\
 $\jpsi$ mass and $\MMS$ selection & 0.4 \\
 Integrated luminosity & 1.4 \\
 $m_{\pp}$ distribution  & 5 \\
 Trigger efficiency & 1 \\
 Branching fractions & 1 \\
 MC statistics & 1 \\
 \hline
 Sum in quadrature & 7.5 \\
 \hline
\end{tabular}
\end{center}
\end{table}

As a validation of our analysis, we measure the $\psp$ cross
section with the same selection criteria. Here 15,444 $\psp$
events survive the selection and the MC-determined detection
efficiency is 5.13\%. This corresponds to \(
\sigma(\psp)=(15.42\pm 0.12\pm 0.89)~\hbox{pb} \) at the
$\Upsilon(4S)$ resonance or \( \Gamma(\psp\to \EE)=(2.54\pm
0.02\pm 0.15)~\hbox{keV}\), where the first error is statistical
and the second systematic. This measurement agrees well with the
world average value of $(2.48\pm 0.06)$~keV~\cite{PDG}. The $\psp$
mass determined from the data indicates the $\pip\pim\ell^+\ell^-$
invariant mass is measured with a precision of $\pm
0.6$~MeV/$c^2$.

An unbinned maximum likelihood fit is applied to the $\pp\LL$ mass
spectrum in Fig.~\ref{mppll_full}. Here the theoretical shape is
multiplied by the efficiency and effective luminosity, which are
functions of the $\pp\LL$ invariant mass. Since there are two
clusters of events in the mass distribution, we fit it with two
coherent Breit-Wigner (BW) resonance functions ($R1$, $R2$)
assuming there is no continuum production of $\EE\to \ppjpsi$. In
the fit, the background term is fixed at the level obtained from a
linear fit to the sideband data, contributions from the $\psp$ and
$\pspp$ resonance tails (added incoherently) are estimated using
world average values for their parameters~\cite{PDG} and fixed,
the widths of the resonances are assumed to be constant. A
three-body decay phase space factor is applied. The MC-determined
mass resolution is less than 5~MeV/$c^2$ over the full mass range.
This is small compared to the widths of the resonances in our
study and is ignored.

\begin{figure}[htbp]
\psfig{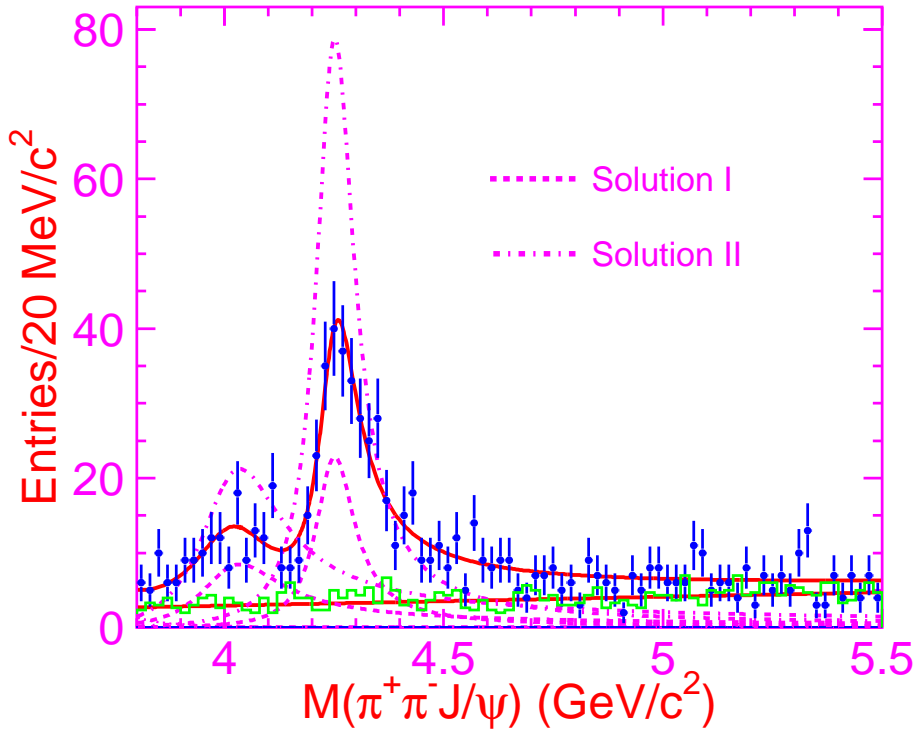} \caption{Fit
to the $\ppjpsi$ mass spectrum with two coherent resonances. The
curves show the best fit and the contribution from each component.
The dashed curves are for solution I, and the dot-dashed curves
for solution II. The histogram shows the scaled sideband
distribution.} \label{coherent_two_res}
\end{figure}

Figure~\ref{coherent_two_res} shows the fit results; there are two
solutions with equally good fit quality. The masses and widths of
the resonances are the same for both solutions; the partial widths
to $\EE$ and the relative phase between them are different (see
Table~\ref{two_sol})~\cite{beebf}. The interference is
constructive for one solution and destructive for the other. The
systematic errors come from the absolute mass scale, the detection
efficiency, the background estimation, the phase space factor, and
the parametrization of the resonances. The quality of the fit
assessed from the binned distribution of
Fig.~\ref{coherent_two_res}, is $\chi^2/ndf=81/78$, corresponding
to a C.L. of 38\%. The statistical significance of the structure
around 4.05~GeV/$c^2$ is estimated to be $7.4\sigma$ from the
change in likelihood value when the BW representing it is removed
from the fit. Although the mass of the first resonance is close to
that of the $\psift$, the fitted width is much wider than its
world average~\cite{PDG} value ($80\pm 10$~MeV/$c^2$). The mass of
the second resonance is higher than that of the $\psifto$. Changes
of resonance parameters that occur when we fit with a coherent
$\psp$ tail, a coherent or incoherent non-resonance term, an
energy-dependent total width, or a cascade two-body phase-space
factor, dominate the systematic errors listed in
Table~\ref{two_sol}; the significance of the $R1$ signal is
greater than $5\sigma$ in all of the fitting scenarios that are
considered. If we use the same functional form as BaBar (a single
BW with an incoherent second-order polynomial background term) we
find $M = 4263\pm 6$~MeV/$c^2$, $\Gamma_{\rm tot} =126\pm
18$~MeV/$c^2$, and $\BR(\ppjpsi)\cdot \Gamma_{\EE} = 9.7\pm
1.1~\hbox{eV}/c^2$, consistent with their
results~\cite{babay4260}.

\begin{table}
\caption{Fit results of the $\ppjpsi$ invariant mass spectrum. The
first errors are statistical and the second systematic. $M$,
$\Gamma_{\rm tot}$, and $\BR\cdot \Gamma_{\EE}$ are the mass (in
MeV/$c^2$), total width (in MeV/$c^2$), product of the branching
fraction to $\ppjpsi$ and the $\EE$ partial width (in eV/$c^2$),
respectively. $\phi$ is the relative phase between the two
resonances (in degrees).}\label{two_sol}
\begin{center}
\renewcommand{\arraystretch}{1.4}
\begin{tabular}{ccc}
  \hline
  Parameters & ~~~Solution I~~~ & ~~~Solution II~~~ \\
  \hline
  $M(R1)$            & \multicolumn{2}{c}{$4008\pm 40^{+114}_{-28}$}  \\
  $\Gamma_{\rm tot}(R1)$   & \multicolumn{2}{c}{$ 226\pm 44\pm 87$}  \\
  $\BR\cdot \Gamma_{\EE}(R1)$
                  & $5.0\pm 1.4^{+6.1}_{-0.9}$ & $12.4\pm 2.4^{+14.8}_{-1.1}$  \\
  $M(R2)$            & \multicolumn{2}{c}{$4247\pm 12^{+17}_{-32}$} \\
  $\Gamma_{\rm tot}(R2)$   & \multicolumn{2}{c}{$ 108\pm 19\pm 10$} \\
  $\BR\cdot \Gamma_{\EE}(R2)$
                  & $6.0\pm 1.2^{+4.7}_{-0.5}$ & $20.6\pm 2.3^{+9.1}_{-1.7}$ \\
  $\phi$          & $12\pm 29^{+7}_{-98}$ & $-111\pm 7^{+28}_{-31}$ \\
  \hline
\end{tabular}
\end{center}
\end{table}

In summary, the $\EE\to\ppjpsi$ cross section is measured for the
CM energy range $\sqrt{s}=3.8$~GeV to 5.5~GeV. There are two
significant enhancements: one near 4.25~GeV, consistent with the
results of Refs.~\cite{babay4260} and~\cite{cleo_y}, and another
near 4.05~GeV, which has not previously been observed. We note
that these enhancements are close to $D^{(*)}\bar{D}^{(*)}$
thresholds, where coupled-channel effects and rescattering may
affect the cross section~\cite{voloshin}. If we nevertheless
represent the cross section using interfering BW terms, a second
term (in addition to the $\y$) substantially improves the fit. In
particular, the lower-mass side of the 4.25~GeV enhancement is
better reproduced. The parameters that are obtained from this
two-term fit do not correspond to those of any of the excited
$\psi$ states currently listed in Refs.~\cite{PDG}
and~\cite{besres}.


We express sincere thanks to H.~Czy$\dot{\hbox{z}}$ for helpful
discussions on the generator. We thank the KEKB group for
excellent operation of the accelerator, the KEK cryogenics group
for efficient solenoid operations, and the KEK computer group and
the NII for valuable computing and Super-SINET network support. We
acknowledge support from MEXT and JSPS (Japan); ARC and DEST
(Australia); NSFC, KIP of CAS, and 100 Talents Program of CAS
(China); DST (India); MOEHRD, KOSEF and KRF (Korea); KBN (Poland);
MES and RFAAE (Russia); ARRS (Slovenia); SNSF (Switzerland); NSC
and MOE (Taiwan); and DOE (USA).


\begin{thebibliography}{**}
\bibitem{beebf} Considering the correlation between
$\BR(\ppjpsi)\cdot \Gamma_{\EE}$ and $\Gamma_{\rm tot}$, we get
\(\BR(\ppjpsi)\cdot \BR(\EE) = (2.2\pm 0.7^{+2.5}_{-0.4})\times
10^{-8}\) and \((5.5\pm 1.1^{+4.2}_{-0.7})\times 10^{-8}\) for
$R1$ and $R2$, respectively, for solution I; and
\(\BR(\ppjpsi)\cdot \BR(\EE) = (5.5\pm 1.5^{+7.0}_{-1.5})\times
10^{-8}\) and \((19.1\pm 2.4^{+8.3}_{-2.3})\times 10^{-8}\) for
$R1$ and $R2$, respectively, for solution II.
\end{thebibliography}
\end{document}